# Thermionic Emission and a Novel Electron Collector
# in a Liquid Helium Environment

J. Fang[1], Anatoly E. Dementyev[1], Jacques Tempere[1,2], and Isaac F. Silvera[1]

[1]Lyman Laboratory of Physics, Harvard University, Cambridge MA 02138, USA.

[2]TFVS, Universiteit Antwerpen, Groenenborgerlaan 171, 2020 Antwerpen, Belgium

We study two techniques to create electrons in a liquid helium environment. One is thermionic emission of tungsten filaments in a low temperature cell in the vapor phase with a superfluid helium film covering all surfaces; the other is operating a glowing filament immersed in bulk liquid helium. We present both the steady state and rapid sweep I-V curves and the electron current yield. These curves, having a negative dynamic resistance region, differ remarkably from those of a vacuum tube filament. A novel low temperature vapor-phase electron collector for which the insulating helium film on the collector surface can be removed is used to measure emission current. We also discuss our achievement of producing multi-electron bubbles (MEBs) in liquid helium by a new method.

## I. INTRODUCTION

The study of free electrons on or in liquid helium started over four decades ago and continues to be an important area of research; many of the fundamental studies are discussed in refs. [1, 2]. There is continued interest [3, 4] in the fascinating properties of electrons on or under the surface. Electrons on the surface of liquid helium form a 2D electron gas and at high enough density can solidify into a lattice, demonstrating Wigner crystallization [5]. There is a 1 eV barrier for electrons to penetrate the surface of liquid helium [6, 7]. Inside of bulk liquid helium, electrons can be in the form of single-electron bubbles (SEBs) with diameters of about 34 Å [8], or multielectron bubbles (MEBs) with macroscopic dimensions, containing hundreds to billions of electrons [9-11][12].

The principal methods of producing large numbers of free electrons have been by thermionic emission (TE) from a hot filament in a helium environment or from a field emission tip (FET) with a submicron radius of curvature that emits above a critical negative voltage [13], usually made by etching down a tungsten wire to a submicron radius of curvature tip. Although both techniques are useful for electron production, the main subject of this paper is thermionic emission in helium at low temperatures. The electron yield is of order picoamps to microamps, leading to large numbers of electrons. The i-V characteristic of a filament in a liquid helium environment is quite different from that of a filament in a vacuum tube. It is interesting to study and understand the reasons for these differences, as well as the operating conditions for TE to produce a desired number of electrons in a low temperature helium environment. The emitting filament can be operated either in the vapor phase of low temperature helium or immersed in liquid helium. It turns out that the same analysis can be used for both the vapor phase operation and liquid immersion operation [14].

In vapor phase operation it is desirable to measure the yield with an electron collector and to remove the electrons from the cell. In a vacuum tube, free electrons produced by thermionic emission from a tungsten filament are removed with a metal collector circuit. In the low temperature environment, the collector is generally covered by a film of helium that presents a barrier for their entry into the metal. In this article, we introduce a method of collecting or removing electrons from such an environment.

## II.   EXPERIMENTAL TECHNIQUES

### A.   Thermionic Emission

Thermionic emission is a well-known, well-studied phenomenon. By heating a metallic wire, electrons populating energy states according to the Fermi distribution can have sufficient energy to escape the potential that holds them in the metal. For temperatures T smaller than the Fermi temperature, the current is given by the Richardson-Dushman equation [15]

$$j = \frac{4\pi mek^2}{h^3}(1-r)T^2 \exp(-\phi/kT),\qquad\qquad (1)$$

where $j = i/S$ is the current density emitted from surface area S, $\phi$ is the work function, r is the reflection coefficient of an electron striking the emitting surface, and k and h are Boltzmann and Planck's constants. Values of $\phi$ for some selected materials are 4.53 eV (tungsten), 2.7 eV (thoriated tungsten), 1.0 eV (cesiated tungsten), and 4.33 eV (copper) [16]. In order to achieve a reasonable emission current in vacuum tubes, high temperatures of order a few thousand degrees K are needed. Tungsten (W) is usually the choice for filaments because of its high melting temperature of 3695 K. By cesiating or thoriating tungsten, the work function can be substantially reduced.

In our vapor-phase measurements we studied pure tungsten and thoriated tungsten wires mounted as shown in Fig. 1A. The thoriated tungsten filament had a diameter of 12 microns, while the diameter of the pure tungsten filament was 25 microns. The lengths were around 10 mm and the room temperature resistances of the order a few ohms.  The filaments were initially attached with silver epoxy to copper posts, mounted in a nylon holder.  Later we electroplated the filaments ends with copper and used solder, resulting in a smaller contact resistance.  The highest filament currents used were around 350 mA.  Currents were generated with a variable dc voltage source; maximum voltages were around 6 volts.  We could also pulse the power with a fast series relay, which allowed us to also apply large bias voltages to our emission/collection electronics.  For use as a low temperature electron source, we found no advantage of one filament wire material over the other, while the pure tungsten was stronger and easier to handle.

For the vapor phase measurements both ends of the filament are anchored to the low temperature bath on copper posts.  At modest power levels (dissipated along the wire) the temperature of the filament peaks at the middle.  Heat loss is due to thermal conduction to the copper posts and heat transfer to the helium gas.  At higher temperatures, radiative (and electron emission) heat losses become important and the central hot region becomes more extended.  In our low temperature experiment in a windowless cryostat we had no means of measuring the temperature of the filament.  However, in a glass cryostat we could observe the glowing filament and estimate the range of temperatures to be 1000-2000 K, depending on the filament current.\

For immersion experiments we used pure tungsten wires. These experiments were carried out in a glass cryostat so that the filament could be viewed through the transparent windows in the silvering of the cryostat with a microscope.  The cryostat was pumped with a high-speed pump and attained temperatures of ~1.3 K and somewhat higher when the filament was operating continuously.  The helium temperature was determined by measuring the vapor pressure with a Wallace and Tiernan pressure gauge.  A condom pressure regulator [17] was used to control the temperature of the helium when studies were in the region of the lambda point of helium.

.B.     **The Electron Collector in the Vapor Phase**

In order to operate an electron collector in the vapor phase experiments, the superfluid helium film must be removed from its surface. The collector, shown in Fig. 1B consists of a foil of gold with an approximately 1 mm$^2$ area attached with insulating epoxy to a resistive heater. The heater is a Dale ruthenium oxide (RuO$_2$) thick film resistor [18] on an alumina substrate, with area $1.6 \times 2.9$ mm$^2$. The surface of the ruthenium oxide film is covered with a protective insulating film. Tungsten wire leads (25 micron diameter), silver epoxied or soldered to the resistor and the collector, were attached to copper mounting terminals pressed into a nylon holder. The resistor is thus weakly linked to the cell so that it is easily heated. The collector was used in the gas phase; it was connected to an electrometer and was not biased so that it operated at virtual ground. Ruthenium oxide resistors can also serve as low temperature thermometers [18]. The 500 ohm (room temperature) resistor was calibrated at helium temperatures as a thermometer. A current of 1 mA was sufficient to heat the resistor a few degrees K above the ambient temperature of about 1.3 K, though the roughly 500 microwatts of power dissipated in the resistor do not measurably affect the temperature of the cell.

The heated resistor evaporates the helium film from its surface at a faster rate than the helium is recondensed from the gas phase. In addition, since the superfluid helium film in the cell tends to flow towards warmer areas, helium flows along the wire leads from the cell surfaces to the resistor, which is warmer than the cell, to replenish the evaporated film. If the heating rate is sufficient, the evaporation will be faster than the replenishment and the collector will become bare of helium, just as trigger bolometers used in the detection of spin-polarized atomic hydrogen [19]. If the temperature is below about 0.8 K, the film is quite robust and for low temperatures there is a clear signature when the film burns off at ~0.8 K. When heated above this temperature there is there is a sharp measurable increase in the temperature of the resistor as the film is burnt off.  With increasing temperature the film becomes more fragile and can be burnt off with small thermal fluxes.  We operated at temperatures of ~1.3 K and higher.

For these vapor phase experiments, both the filament and collector were mounted in a cylindrical brass cell with a free volume of about 10 cm$^3$, but with a large surface area as the cell contained a large plastic filler to reduce the volume. The cell itself was filled with enough helium to easily saturate all surfaces with a film.  There was a mask between the filament and the collector so that there was no line-of-sight path between the two and the filament holder was facing down at the helium covered plastic filler to suppress the possibility of burning the helium film off of the metallic surfaces of the cell walls, which then act as a leakage path for the electrons.  The sealed cell was immersed in a pumped liquid helium bath so that heat removal was efficient.

The collector current was detected with a Keithley 617 electrometer in the current mode. If the collector, sitting in the vapor phase, is to be used to drain off electrons from the cell in a controlled way, it is important to estimate the time constant for the current discharge of a cell filled with N electrons.  Preferably, this time constant should be relatively long, so that the charge in the cell can be controlled by opening or closing a switch to a conducting wire connected to a charge drain.  To estimate the time constant we first assume that the free electrons are uniformly distributed throughout the volume V of the cell and in thermal equilibrium. Because the density is low and the temperature relatively high, the distribution can be described by a Maxwell-Boltzmann distribution, rather than a Fermi distribution.  In this case, the average flux of electrons hitting the surface of the collector area A and entering the metal is

$$\frac{dN}{dt} = -\frac{1}{4}n\bar{v}sA \qquad (2)$$

where $n = N/V$ is the gas phase density of electrons, $\bar{v} = (8kT/\pi m)^{1/2}$ is the average velocity of the electrons with mass m, and s is the probability that an electron that hits the collector surface enters the metal. Solving Eq. (2), one finds that the cell discharges as $N(t) = N_0 \exp(-t/\tau)$ with a time constant

$$\tau = 4V / sA\bar{v} . \tag{3}$$

For purposes of estimation, we take s to be 0.5. With $V \approx 10$ cm$^3$, $A \approx 0.01$ cm$^2$, T=1.3 K, and $\bar{v} = 7.1 \times 10^5$ cm/s; we find $\tau = 2.9 \times 10^{-3}$ s.

Since there is a binding energy of the electrons to the helium surface [20], in equilibrium electrons will distribute into volume and surface states, and thus there is an electron surface density. The binding energy of an electron on a helium film on a metal surface can be substantially larger than on bulk helium due to the image charge of the electron in the metal. (The image charge above a dielectric is reduced by the factor $(\varepsilon - 1)/(\varepsilon + 1)$ where $\varepsilon$ is the dielectric constant of the insulator). The above calculation only considers the discharge of the bulk electrons. Since the electrons are charged they will repel and drive each other to the cell surfaces, except when they have sufficient energy. Thus, in equilibrium we expect the discharge to be slower as electrons flow mainly in surface states, rather than in the volume.

III.    **Experimental Results**

**A.  The Vapor Phase**

If the filament is operating in a vacuum, as the voltage increases the current increases monotonically; the filament warms due to ohmic heating and the resistance rises as the temperature rises. At a high enough i-V point the filament becomes very hot, glows and emits electrons. The behavior is very different in the vapor phase as shown in Fig. 2. Most low temperature thermionic emission sources are used in a pulsed mode, which is sufficient to produce enough electrons without excess heating or change in cell temperature [21]. In this study we used a cell with efficient heat removal to maintain a low temperature in the presence of large power dissipation so that we could also operate in continuous mode to understand the "fly-back" or reduction of current with increasing voltage, as seen in Fig. 2. We first discuss the steady state i-V curve of a filament wire, using a <u>constant voltage source</u>, shown in Fig. 2 (continuous curve), studied earlier by Silvera and Tempere. As the voltage increases the current increases until a maximum is reached and the current falls back in a negative dynamic resistance region. The reduced current is due to an increase in the absolute resistance of the filament, while V is in a constant voltage mode. Further increase of the voltage leads to a relatively small increase in the current as the resistance has made a transition to a high value branch. Thus, we consider the filament to have a high resistance (upper) branch and a low resistance (lower) branch. It is on the upper branch that the filament is hot and emits a significant number of electrons. In reducing the voltage, the curve is retraced with no hysteresis. Note that if a constant current source were used, then at the critical voltage $V_c$ (point 1 in Fig. 2, where the slope of the i-V curve is infinite) the voltage would jump vertically to operating at a higher voltage. Due to the increased resistance the ohmic heating increases at this value of current and the filament temperature becomes much higher than when operating with a constant voltage source where the power dissipation reduces rather than increases by a large jump. In the constant current mode one must be careful with further increasing the current as the filament may be near its burn-out current. Thus, it is prudent and safer to operate the filament in a constant voltage mode, rather than a constant current mode.

If the voltage is rapidly increased above or below the critical voltage, then the i-V curve shows substantial hysteresis. This is due to the thermal time constant for the temperature of the filament and its environment to change. In Fig. 2 we see that starting on the lower branch, if the voltage is rapidly increased above $V_c$ from point 1 to point 2, then the curve momentarily traces up the low resistance branch away from the steady state equilibrium curve and then jumps to the high-temperature high-resistance branch (from point 2 to point 3). Rapidly reducing the voltage below $V_c$, the curve follows the high resistance branch (from point 3 to point 4) and then jumps to the low branch (from point 4 to point 1). We did not perform a systematic study of the time constant for recovery to the steady state curve, as this depended on the particular conditions (temperature, voltage jump, etc). In the example shown, if the voltage was abruptly changed (~1 ms), the time for recovery to the steady state curve was approximately 1s. For obvious reasons the reduction of current in pulsed operation is called the fly-back.

In Fig. 3 we show the collector current as a function of the steady state filament current when energizing the tungsten filament. The two curves correspond to changing the voltage to move up the upper branch or down the lower branch from the region of $V_c$. The current yield is of the order of a few picoamps. The film burner on the collector was operating for these measurements. We expected a stronger dependence of the current according to Eq. 1. It was suspected that the hot filament locally burns off the He film at the inner wall of the cell (which was grounded), giving a parallel current path. This was confirmed later in a different apparatus. In addition the electrons may tunnel through the image-charge thinned helium film covering metallic surfaces in the cell, although this has been observed to be a much slower process [22]. Thus, in our experiments, our collector is sampling the emission current, rather than measuring the total emitted current.

Our test of the effectiveness of the new low temperature collector gave mixed results. Under certain conditions we observed that activating or deactivating the resistor-heater caused the collector current to flow or stop; however, for some tests we found that current flowed to the collector even when the resistor was deactivated. Our explanation is the following: when the filament is operating on the upper branch it is hot and glowing brightly. The collector can be heated by radiation, hot electrons, or energetic helium atoms, scattering off the filament that can thin or burn off the film. For cell temperatures below about 1 K where the helium film is robust, the collector should operate reliably; however, in our test apparatus the temperatures were 1.3 K and higher and the film becomes fragile. When operating the filament on the upper branch, it dissipates up to ~400 mW which is high for a low temperature environment and leads to gradual warming of the helium bath from 1.3 K to as high as ~1.7 K. At the higher temperatures the film can easily be burned off of the collector even without heating of the resistor.

The behavior of electrons in environments such as ours seems to be complex and the literature is filled with seemingly contradictory results where either the electrons (in the presence of helium covered metallic walls) have a moderate lifetime [23], or disappear more rapidly after being produced by a filament or an FET [6, 22, 24]. Here we share some of our experiences to overcome the problem of loss of electrons to surfaces with or without helium coverage. We have placed a copper "holding plate" with a bias of several hundred volts (positive) in the cell and manipulate liquid helium so that there is a ~1 mm thick liquid helium layer above the plate. The electron source in the vapor phase is then activated for a period of time. Electrons accumulate above the holding plate on the helium surface due to the positive bias on the plate and are then very stable as a two-dimensional electron gas. After the source is turned off and the cell equilibrates, so that all surfaces in the vapor phase are covered with a helium film, we turn off the holding plate bias and electrons are released. This method has been successful in producing electrons with a longer lifetime in the vapor phase or on the exposed helium covered surfaces.

Accurately measuring low-level currents is challenging. In our collector we had a leakage current of a few picoamps that we could not suppress. It is not unusual to have such a leakage current across the leads, as our tests with the Keithley electrometer showed. More unusual was that the leakage current depended on the magnitude and polarity of the current heating the ruthenium oxide resistor. Since this can change the local electric fields in the vicinity of the collector, we believe that the leakage current was due to a large (gigaohm) resistance between the resistor and the gold collector that was attached with Emerson and Cumming 2850 FT epoxy. In any case the leakage currents were measured to correct the collector current readings.

We tried to measure the time constant for discharging the cell after it was loaded with electrons. This proved to be difficult due to noise, low-level signals, and the response time of our measuring apparatus. To amplify the effect we ran the filament for 1 to 2 minutes with an open switch in the line to the electrometer so that the charge would build up. However, closing the simple switch or a mercury-wetted switch created a noise pulse that masked the electron signal. We finally found that we could just disconnect and reconnect the coax cable to the ammeter by hand to reduce the noise pulse to a level below the electron signal, but the approximate 0.1 s response time of our measurement apparatus was limiting, so we could only set this as an upper bound.

The picoamp level is actually quite high for certain experiments, corresponding to a flow of $6.24 \times 10^6$ electrons/s. Lower yields can be achieved if the filament is pulsed on and off with a voltage pulse such that the time of the current can be controlled, or alternately, can be achieved by working on the lower branch of the filament near $V_c$ (see Fig. 3). Pulsing the filament on and off is an important method if the cell has limited cooling power. When operating in a different cell with poor heat removal, as the temperature rose, the gas pressure rose and no fly-back was observed on the i-V curve, i.e. the filament failed to jump to the high R branch. The reason for this was that the higher-pressure helium gas can continue to efficiently remove heat from the filament in a convective mode, as shall be discussed ahead. Pulsed operation has the advantage that the average heating is minimized and the charge can be controlled by the number of pulses.

The heat dissipated in the larger diameter tungsten filaments is substantial.  We have also used filaments with 10 and 5 micron diameters, as an analysis indicated that these would result in much less heating [14].  The 5 micron diameter filaments dissipated about 10-20 mW at the critical voltage.

**B.        Immersion in Liquid Helium**

Spangler and Hereford [25] studied the electron emission and current from a tungsten filament immersed in liquid helium.  They found that above a certain current a vapor sheath formed around the filament, insulating it from the liquid and it would glow at temperatures up to a few thousand K, emitting large electron currents to a cylindrical collector in the helium around the filament.  Okuda et al [26] measured the sheath diameter to be about ~100 microns at formation.  Date et al [27] also observed a negative resistance region in the i-V curve as we have seen in the vapor phase.  In our experiment we have suspended a 25 micron diameter tungsten filament about 10 mm in length between electrodes in liquid helium.  A collector, simply a silver-coated copper sheet of area ~1 cm$^2$ was situated above the filament in the liquid helium.

The following discussion is for liquid helium temperature, T~1.5 K.  First we have reexamined the i-V curves and show our results in Fig. 4.  In this case the current reduction (fly-back) is extremely sharp and occurs when the vapor sheath is formed, often with an audible hiss at a frequency of a few hundred Hz.  Visible oscillations of the filament, perhaps at tens of Hz, were visually observed with a microscope.  With increasing voltage (hotter) the filament began to glow brighter and brighter, and the sheath expanded.  At lower power the surface of the sheath was quite stable. As the power was increased the sheath become somewhat chaotic and oscillated, but it was tethered to the filament, never breaking loose as a bubble filled with electrons, which was our objective.  Most of our measurements were made at a depth of 10 to 15 cm below the surface of the helium; results were somewhat dependent on the pressure head of the helium [27].

In order to study the electron emission in the sheath, we placed a small copper wire close to the filament so that it would penetrate the sheath, to act as a collector. The filament could be biased with a negative voltage relative to the copper wire. The electron current collected in the copper wire was quite low but increased rapidly to over 50 microamps as the negative filament voltage was increased, as shown in Fig. 5. Space charge around the filament holds the cloud of electrons close to the emitting surface. As the negative bias of the filament is increased the cloud expands to better overlap the collector, which we interpret as the reason for the increased collector current.

We now discuss how the sheath is formed. As the filament current is increased from zero the heat generated is transported from the surface of the filament directly into the superfluid. Eventually as the heat flux increases, the superfluid cannot carry the thermal load and a vapor sheath forms around the filament. The heat transport mechanism goes over to one of heat transport into the vapor and then into the surrounding liquid with a much reduced heat transfer. The filament gets hotter, its resistance increases, and the current falls back since we use a constant voltage circuit. The heat transfer mechanism becomes one of helium atoms evaporating from the hot surface of the filament; the thermal flux is adsorbed on the surface of the sheath. The hot atoms evaporated from the surface of the filament transfer momentum to the surface of the sheath (the sticking coefficient of helium atoms on a helium surface is ~1) to maintain the equilibrium surface diameter, while cold atoms evaporating from the liquid surface of the sheath adsorb on the filament surface, thermalize and are re-evaporated to continue the cycle. Thus, when the filament is hot enough the sheath is filled with helium vapor and electrons. The electrons tend to form around the filament as space charge due to their image charges in the filament. The heat removal is by convective superfluid flow to the surface of the sheath; the surface oscillates, but visible bubbles are not released.

It is also possible that the emitted electrons help maintain the diameter of the sheath. To test this we increased the bias on the filament to about -150 Volts and it did not affect the diameter of the sheath. However, with a much higher potential (-2 KV) the sheath became larger so that at this voltage level the electrons play a role in the sheath geometry.

## IV.    An application of Thermionic Emission: Creating MEBs

In Spangler and Hereford's [25] measurement at T=1.3 K, a 5 micron diameter wire was surrounded by a 2 cm diameter cylindrical anode with a bias voltage up to -2.6 KV. They observed very low electron currents at low bias and currents up to 0.5 mA at the highest bias. This was attributed to single-electron bubble flow from the cathode to the anode and has been modeled by Hori et al [28]. We have measured the current from the immersed filament to our immersed silver-coated collector (~3 cm above the filament) as a function of temperature of the helium bath while our submerged filament was glowing, shown in Fig.6. We did not apply a bias voltage between the filament and collector and for low temperatures (~1.5 K) the current was negligible.   At the lambda point of helium the collector current rises rapidly to a peak value and then slowly decays with further increase in temperature.  As the lambda point is approached the convective heat removal of superfluid helium is reduced to zero and the heat removal at the sheath boundary is greatly diminished.  We observed torrents of bubbles being created at the filament, rapidly rising in the helium bath due to buoyancy.  These bubbles were produced all along the filament, but in the hot glowing region, the filament is also emitting electrons.  We believe that the explanation for the electron current at the collector is that bubbles formed in the hot glowing region are filled not only with helium gas, but also electrons and form multielectron bubbles [4] that rise up to the collector.  When a bubble contacts the collector surface there is a thin helium film between the electrons and the metal, and the electrons can tunnel through. We attempted to confirm that the bubbles contained electrons by placing a strong electric field transverse to the path of the bubbles as they rose and looking for a deflection, but we were unable to determine a deflection as the torrent of neutral bubbles, mostly from the non-glowing regions of the filament where electrons are not emitted, masked the observation. The bubbles as observed by eye were quite large (~1-10 mm) and expanded as they rose. In an attempt to improve the visibility of charged bubbles we pulsed the voltage on for tens of milliseconds to create a burst of bubbles, but the putative MEBs were still masked by the more numerous neutral bubbles.

We then used a fast current amplifier, rather than the electrometer, to examine the charge arrival at the collector, shown in Fig. 7. We observed sharp narrow (width ~ 1 ms) current pulses. Integrating the area under say, the first current pulse, yields the number of electrons as $\sim 1.5 \times 10^7$. This is strong evidence that some of the bubbles are carrying electrons, i.e. they are MEBs rising in the helium. If the electrons were transported as SEBs then we would expect a distribution of very small amplitude short current pulses. These may also exist but if so, they are buried in the noise level of Fig. 7. Thus, we have created MEBs with this technique, but only at or above lambda point of helium. The SEBs of Spengler and Hereford were created below the lambda point with a large electric field gradient. In our case we find that below the lambda point large visible bubbles or MEBs could not be released from the sheath.

## V.    A Model for the Electro-thermal Properties of a Filament

The unusual steady state i-V curve of the filament in the helium environment differs from the behavior of a tungsten filament in a vacuum tube. In the former, the slope has a negative dynamic resistance region, whereas in the latter, the resistance increases gradually with current or temperature, with an almost linear i-V curve. The main difference in these two cases is that in the vacuum tube there is no exchange gas that can remove heat from the filament. Thus, we assume that the i-V characteristic in our study is strongly influenced by the helium gas, film, or liquid. Helium film flow, with the filament in the gas phase, turns out to be unimportant for the understanding of the TE at low temperature. At lower current or lower temperature of the filament, the gas cooling is convective, since the cold helium gas flows towards the warmer region of the filament, rapidly expands more than a hundred fold as it warms up and convects heat away. At the critical voltage (current), as the environment becomes warm, convection ceases and the thermal conduction goes over to a chaotic diffusive mode [29], and the heat transfer is reduced. In the crossover region a small increment in voltage leads to a disproportional increase in heating of the filament and thus a larger increase in the filament resistance, so that the current drops, leading to the negative resistance region. This behavior has been reproduced theoretically by Silvera and Tempere [14]; their results are shown in the inset of Fig. 4. They solved the differential equation for the heat flow in a filament to determine its temperature profile and resistance. The temperature dependence of the Nusselt number, the ratio of the convective to the conductive heat transport, is shown in the inset. By varying the temperature dependence, the behavior in the vapor phase (Fig. 2) or sharper changes for the immersed filament with the sheath (Fig. 4) can be reproduced.

The response of the filament i-V characteristic to an instantaneous increase in voltage, as in Fig. 2, is consistent with this explanation. In this case, initial cooling of the filament is convective and relaxes to non-convective heat transport that results in the fly-back to the steady state curve. This hypothesis was confirmed in an experiment in which the filament was transferred to an experimental cell that had poor thermal contact to a pumped helium bath, compared to the test cell. In this experiment, a negative resistance region was never observed. Our explanation is that with a power dissipation of order 1 Watt in the filament, the experimental cell rapidly heated up from 1.2 K to about 4 K and the helium pressure rapidly rose from of order 1 torr to several hundred torr so that the heat removal from the filament by helium gas was efficient, thus no relaxation to non-convective heat transport and no fly-back. However, when much of the helium was removed so that there was approximately 20 torr pressure at 4.2 K, sufficient to form a film, and then the cell was cooled to about 1.2 K, the negative resistance behavior was recovered. In this case even though the cell warmed there was limited density of helium gas and the heat removal from the filament by helium gas was not efficient.

## VI.    Conclusion

We have shown how to characterize a filament for low temperature electron emission, its current yield, and how to measure the current with a novel collector. This should be useful for studies of electrons in a liquid helium environment. We have also expanded the study of a filament immersed in helium and shown that the sheath that is formed around a hot filament is filled with electrons and that if this bubbles are released they can be multielectron bubbles.

## VII.    Acknowledgements

J. Huang and J. Sommers, as well as H. Kim, contributed to early studies and their efforts are acknowledged. IFS thank Mike Grimes for providing the thoriated tungsten wire some time ago, with comments on its pulsed utilization. This research was supported by DoE grant No. DE-FG02-ER45978, and by the FWO-V Projects Nos. G.0356.06, G.0115.06, G.0180.09N, G.0370.09N.

**Figure Captions**

Figure 1.  Cross sectional views of thermionic source and electron collector.  On the left is the
filament connected to copper terminals in a nylon holder.  Silver epoxy or solder was
used for the contacts (on copper plated wire ends); on the right the electron collector,
described in the text.

Figure 2.  The current-voltage curves for:  A. thoriated-tungsten, and B. pure tungsten filaments
operated with a voltage source.  The steady-state equilibrium curves and rapid change
hysteretic curves are described in the text.  For the hysteretic curves the voltage is
rapidly increased from point 1 to 2 (increase time ~1 ms) and after a delay the current
flies back to point 3 on the steady state curve.  Reducing the voltage from point 3 to 4 is
followed in time by a return to the original point 1.  Slight offsets from the steady state
curves are due to changing temperature conditions.  Temperatures were in the range 1.4-
1.5 K.

Figure 3.  The current yield on the collector, from the tungsten-filament operating on the upper and
lower resistance branches shown in Fig. 2.  The positive current values for the lower
branch are consistent with zero when drift in the leakage current is taken into account.
The statistical error on the collector current measurements (not shown) was of order a
few percent, but there were larger long-term drifts of order a few tenths of a picoamps,
attributed to changing temperature of the cell during measurement.

Figure 4.  The i-V characteristic for a 10 micron diameter tungsten filament immersed in liquid
helium bath at a temperature ~1.5 K.  Note the very sharp current fly-back at the critical
voltage compared to that in Fig. 2.  The inset shows a theoretical analysis of the i-V
characteristic from ref. [14], discussed in the text.

Figure 5. The electron current collected on the small copper wire in the sheath around a glowing
filament immersed in liquid helium as a function of the bias voltage on the filament at a
bath temperature of ~1.5 K.

Figure 6. The electron current of a collector immersed in helium, ~3 cm above a glowing filament
as a function of the temperature determined by the vapor pressure above the liquid
helium. The filament was not biased.

Fig. 7. Current pulses as a function of time for bubbles rising in helium at or above the lambda
point. The upper curve is the amplified current at the collector (the current scale on the
left is only for the collector). The lower curve shows an ~200 ms long pulse of current
through the filament, which was not biased with respect to the collector (~3 cm above
the filament). In this example, approximately 90 ms are required to form the sheath,
release bubbles, and for the bubbles to arrive at the collector. We see two pulses
corresponding to two MEBs. The inset is an expansion of the scale for one of the
current pulses. Integrating the area under this curve yields a charge of ~2.4 pC or
$1.5 \times 10^7$ electrons.

A.

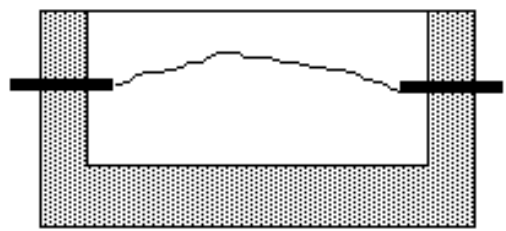

Filament holder

B.

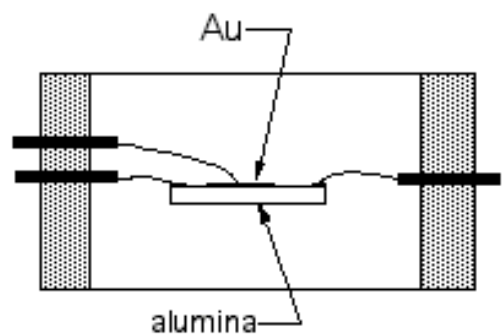

Au

alumina

Collector holder

Fig. 1

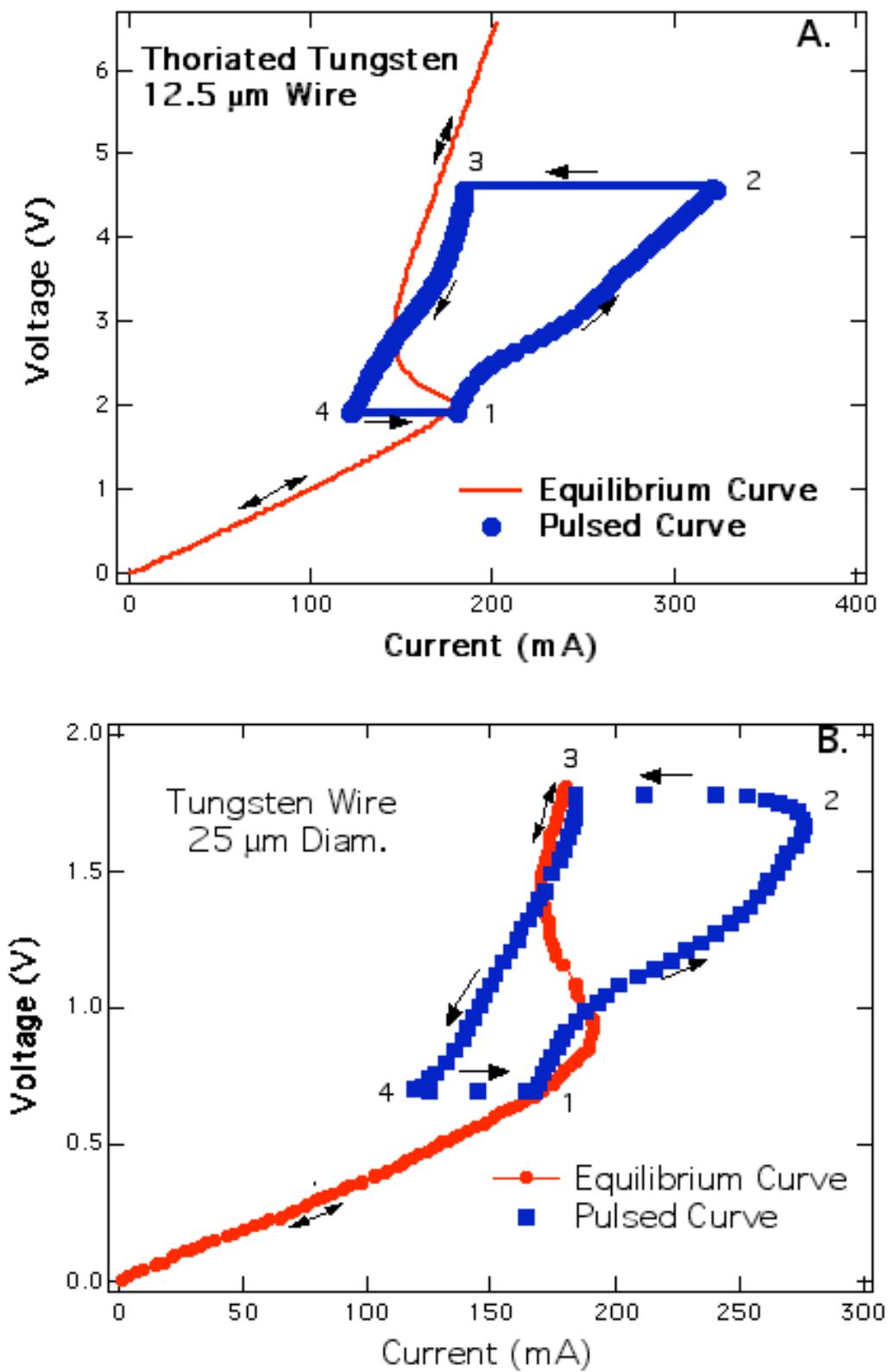

Fig. 2

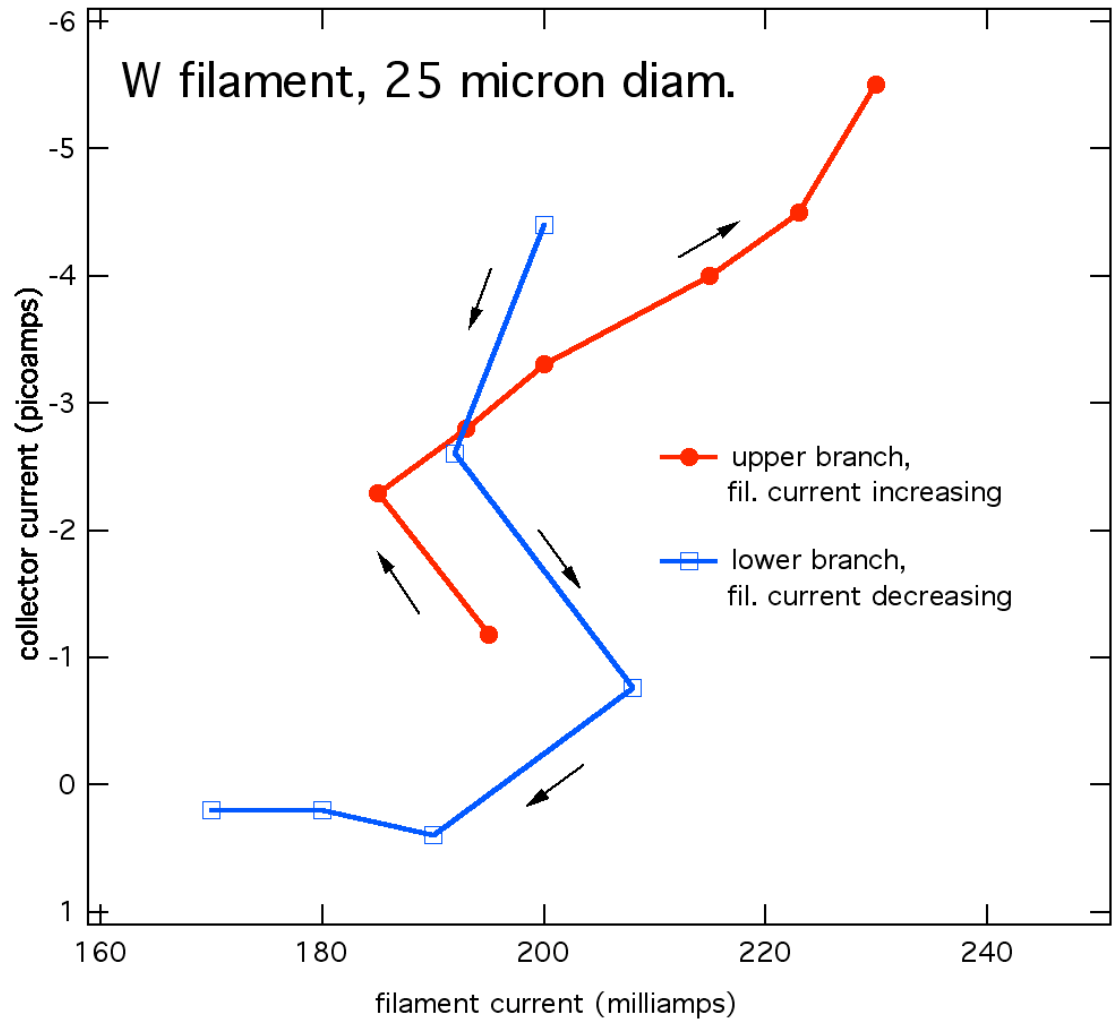

Fig. 3.

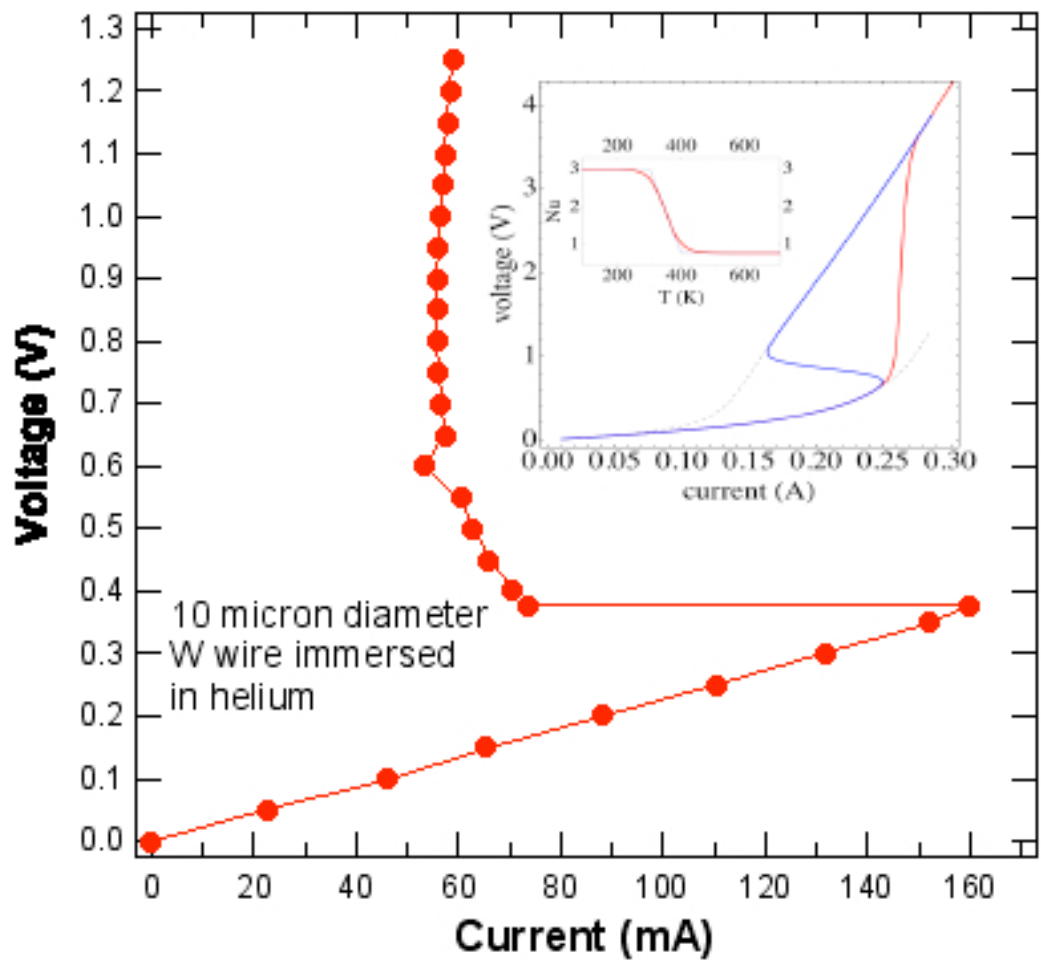

10 micron diameter
W wire immersed
in helium

Fig. 4

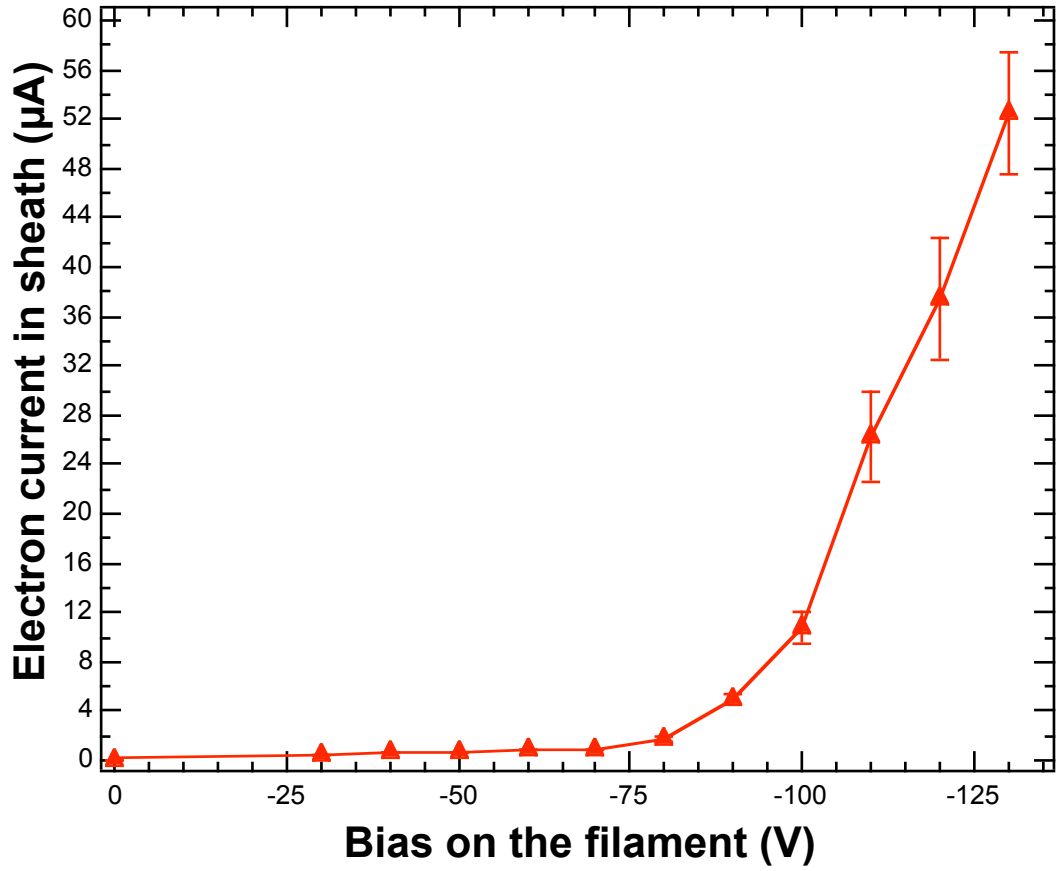

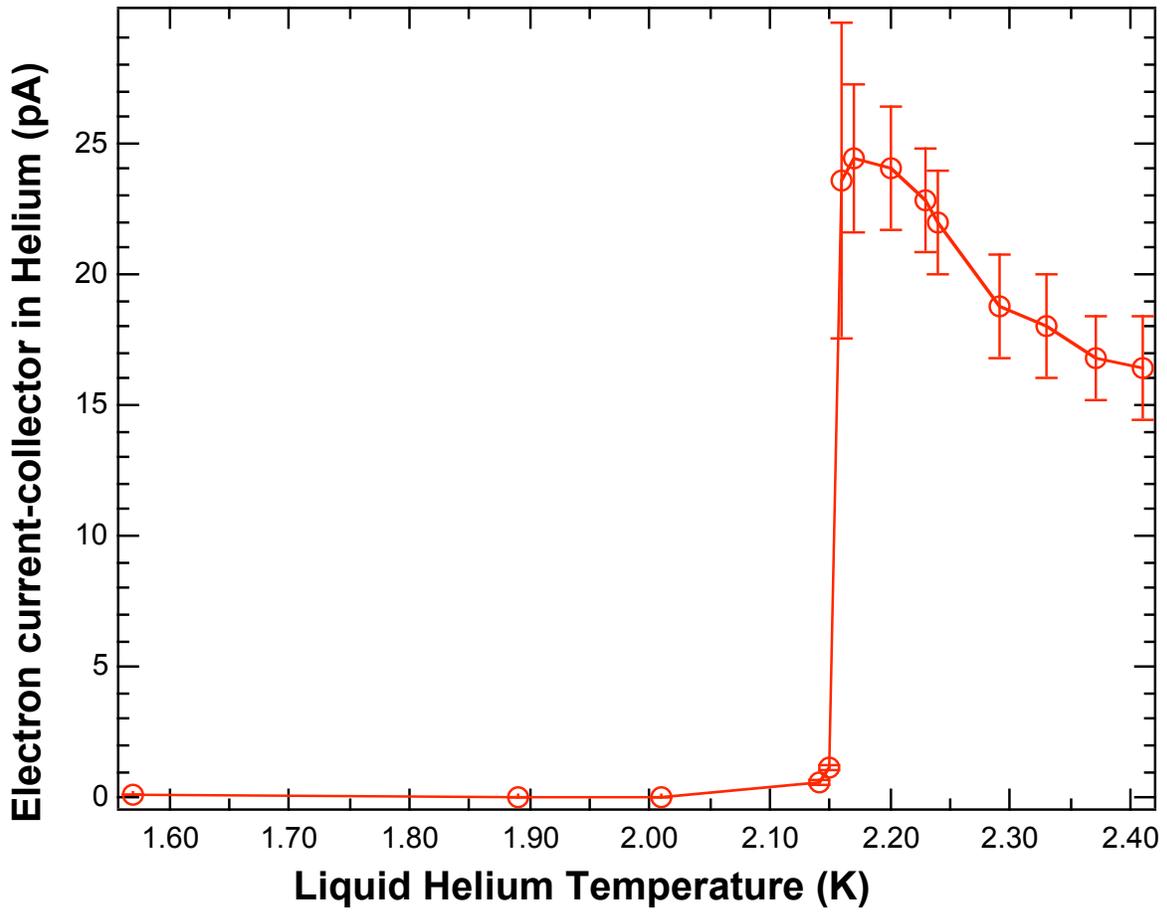

Fig. 6

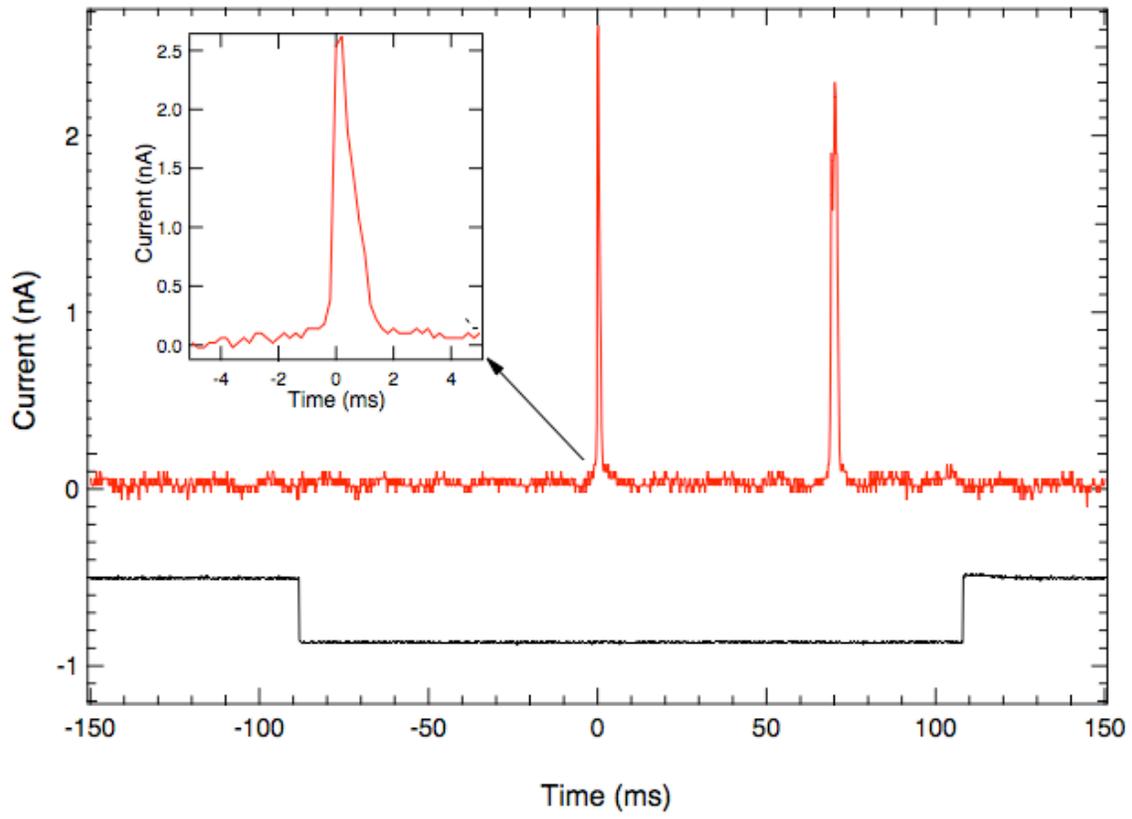

Fig. 7